\documentclass[12pt,preprint]{aastex}

\newcommand{\eqb}{\begin{eqnarray}}
\newcommand{\eqe}{\end{eqnarray}}
\newcommand{\diff}{{\textrm{d}}}
\newcommand{\ellb}{\ell_{\rm B}}
\newcommand{\elli}{\ell_{\rm inj}}

\shorttitle{Automatic quenching of high energy $\gamma$-rays}
\shortauthors{Stawarz \& Kirk}

\begin{document}

\title{Automatic quenching of high energy $\gamma$-ray sources by synchrotron photons}

\author{\L ukasz Stawarz\altaffilmark{1,2} and John G. Kirk\altaffilmark{3}} 
\affil{$^1$KIPAC/SLAC, Stanford University, Stanford, CA 94305\\
$^3$Max-Planck-Institut f\"ur Kernphysik, Saupfercheckweg 1, 69117 Heidelberg, Germany} 
\altaffiltext{2}{Also at OAUJ, ul. Orla 171, 30-244 Krak\'ow, Poland. E-mail: \texttt{stawarz@slac.stanford.edu}}

\begin{abstract}
We investigate a magnetized plasma in which injected high energy gamma-rays annihilate on a soft photon field, that is provided by the synchrotron radiation of the created pairs. For a very wide range of magnetic fields, this process involves gamma-rays between $0.3$\,GeV and $30$\,TeV. We derive a simple dynamical system for this process, analyze its stability to runaway production of soft photons and paris, and find conditions for it to automatically quench by reaching a steady state with an optical depth to photon-photon annihilation larger than unity.  We discuss applications to broad-band $\gamma$-ray emitters, in particular supermassive black holes. Automatic quenching limits the gamma-ray luminosity of these objects and predicts substantial pair loading of the jets of less active sources.
\end{abstract}

\keywords{radiation mechanisms: non-thermal --- gamma-rays: theory --- galaxies: active}

\section{Introduction}

Astrophysical sources of high energy X-ray and $\gamma$-ray radiation are often compact enough for significant photon-photon annihilation \citep{jel66,bonomettorees71,her74,gui83,kaz84}, leading to electromagnetic cascades. In many situations, the cascades can be assumed to be linear, in the sense that the density of target photons is fixed. This is particularly convenient for a Monte-Carlo computation of the spectrum \citep[e.g.,][]{aha85,pro86}. Nonlinearity arises when photons produced by the cascading electron-positron pairs are themselves targets in the photon-photon pair-production process. This has been considered mostly in connection with compact X-ray sources \citep{sve87}, where photons that are inverse-Compton scattered by the nonthermal pairs form the target population.   

Using simplified kinetic equations, \citet{zdz85} studied the steady state radiative spectra of non-thermal electrons in active galactic nuclei subjected to inverse-Compton and synchrotron energy losses, taking photon-photon annihilation and the resulting production of a first generation of secondary electron-positron pairs into account self-consistently. A similar discussion was presented by \citet{fab86} and \citet{lig87}, who investigated the spectra of such sources for a wide range of parameters and included the effects of thermal pairs. Facilitated by the dramatic improvement in computing resources, algorithms for the computation of time-dependent solutions of the kinetic equations describing these processes have been developed by several groups \citep[e.g.,][]{coppi92,mastichiadiskirk95,sternetal95} and are routinely used in source modeling \citep{konopelkoetal99,krawczynskietal02}. Recently, however, attention has focused on intrinsically nonlinear effects. For example, \citet{kirkmastichiadis92} studied the conditions under which ultrarelativistic protons initiate runaway pair-production. For active galactic nuclei, the parameters are extreme, but gamma-ray bursts provide a more promising environment \citep{kazanasetal02,mastichiadiskazanas06}. \citet{stern03} and \citet{sternpoutanen06} have found a strong feedback mechanism that may mediate relativistic shocks, or, in relativistic jets, breed high energy $\gamma$-rays.

In this {\em Letter} we analyze the conditions under which a magnetized system, with an embedded source of photons of energy $\varepsilon \equiv \epsilon \, m_{\rm e} c^2\gg1$ can sustain pair production autonomously. The nonlinear process that makes this possible is annihilation on soft photons with energies $\varepsilon_0 \equiv \epsilon_0 \, m_{\rm e} c^2\ll 1$, produced as the synchrotron emission of the created electron-positron pairs. To describe the dynamics of this process, we simplify the description to a system of three, first-order, ordinary differential equations for the evolution on timescales larger than the light-crossing time of the source, of the populations of hard photons, soft photons and electrons. To this end, we assume an isotropic distribution of the photon fields and relativistic pairs in the source rest frame. The described process operates when the magnetic field $B$ is such that
\eqb
B &=& 6 \, B_{\rm cr} \, \epsilon^{-3} \quad \,
\label{resonance}
\eqe
where $B_{\rm cr} = \, m_{\rm e}^2 c^3 /\hbar \, e \approx 4.4 \times 10^{13}$\,G. This implies $B = 265 \, \left(\epsilon / 10^6\right)^{-3}$\,$\mu$G $= 35 \, \left(\varepsilon / {\rm TeV}\right)^{-3}$\,$\mu$G. Such systems are important in high energy astrophysics, because for an extremely wide range of magnetic field strengths ($B = 10^{-9}$\,G$-10^6$\,G) Eq.~(\ref{resonance}) is satisfied for photon energies between $\varepsilon \approx 0.3$\,GeV and $30$\,TeV, that are accessible to space or ground based $\gamma$-ray observatories.

The reduction of a similar complex system of relativistic particles and photons to a dynamical system of two ordinary differential equations was discussed in an early paper by \citet{sternsvensson91}, who found limit-cycle solutions. In principle, the dynamical system we derive is of sufficiently high order (third) to display much richer \lq\lq chaotic\rq\rq\ behavior. However, the physical processes we discuss are strongly dissipative; in the system studied, we find no physically acceptable oscillatory or quasi-oscillatory solutions and no chaotic regime. Nevertheless, the system has interesting stable solutions with implications for the $\gamma$-ray spectrum of supermassive black holes, and the electron-positron pair loading of their jets.

\section{Equations}

We restrict our analysis to the high energy and soft photon populations that interact with ultrarelativistic pairs (henceforth \lq\lq electrons\rq\rq) of characteristic Lorentz factor $\gamma$. In the systems of interest, the radiative cooling time scale of these electrons is much shorter than the light-crossing time of the emitting volume, and we assume that no other process such as acceleration, or interaction with background particles or waves competes with the cooling. Although our formulation does not enable us to extract detailed spectral information, we discuss the implications for the case of a broad-band spectrum of high energy photons.

We denote the total number of high energy photons, soft photons, and the created electrons in our source volume $\mathcal{V}$ by $N$, $N_0$ and $N_{\rm e}$, and define the effective logarithmic number densities as $\mathcal{N}(\epsilon)=N/\epsilon\mathcal{V}$, $\mathcal{N}_0(\epsilon_0)=N_0/\epsilon_0\mathcal{V}$, $\mathcal{N}_{\rm e}(\gamma)=N_{\rm e}/\gamma\mathcal{V}$. The kinetic equations are then
\eqb
{\diff \, \mathcal{N}(\epsilon) \over \diff t} &=& \dot{\mathcal{N}}^{\gamma \gamma}(\epsilon) + \dot{\mathcal{N}}^{\rm esc}(\epsilon) + \mathcal{Q}(\epsilon) \quad ,
\nonumber\\
{\diff \, \mathcal{N}_0(\epsilon_0) \over \diff t} &=& \dot{\mathcal{N}}_{0}^{\rm esc}(\epsilon_0) + \dot{\mathcal{N}}_{0}^{\rm syn}(\epsilon_0) \quad ,
\nonumber\\
{\diff \, \mathcal{N}_{\rm e}(\gamma) \over \diff t} &=& \dot{\mathcal{N}}_{\rm e}^{\gamma \gamma}(\gamma) + \dot{\mathcal{N}}_{\rm e}^{\rm syn}(\gamma) + \dot{\mathcal{N}}_{\rm e}^{\rm ic}(\gamma)
\label{kinetic}
\eqe
\citep[see][]{zdz85,fab86}. The terms on the right-hand sides of the 
Eq.~(\ref{kinetic}) are: (i) the injection rate $\mathcal{Q}(\epsilon) = L_{\rm inj} / (\epsilon^2 \, m_{\rm e} c^2 \, \mathcal{V})$ of high energy photons into the emission volume $\mathcal{V} = \pi R^3$, where $L_{\rm inj}$ is the high energy luminosity intrinsic to the source. The annihilation rate of these photons $\dot{\mathcal{N}}^{\gamma \gamma}(\epsilon) = - c \, \alpha_{\gamma\gamma} \, \mathcal{N}(\epsilon)$, with $\alpha_{\gamma\gamma} = (2/3) \, \sigma_{\rm T} \, \epsilon^{-1} \, \mathcal{N}_0(\epsilon_0)$, as follows from the approximate expression for the photon-photon annihilation cross section, $\sigma_{\gamma \gamma}(\epsilon, \epsilon_0) = (1/3) \, \sigma_{\rm T} \, \epsilon_0 \, \delta\left[\epsilon_0 - (2 / \epsilon)\right]$ \citep{zdz85}, and their escape rate $\dot{\mathcal{N}}^{\rm esc}(\epsilon) = - \mathcal{N}(\epsilon) / t_{\rm esc}$ with $t_{\rm esc} = R/c$ (this defines the effective radius $R$).
(ii) The soft photon production rate by synchrotron emission $\epsilon_0\dot{\mathcal{N}}_0^{\rm syn}(\epsilon_0) = 4 \pi \, j_{\epsilon_0} / (m_{\rm e} c^2)$, with the synchrotron emissivity $\epsilon_0 j_{\epsilon_0} = \gamma \, \mathcal{N}_{\rm e}(\gamma) \, |\dot{\gamma}|_{\rm syn} \, m_{\rm e} c^2 / (8 \pi)$. the synchrotron energy losses term $|\dot{\gamma}|_{\rm syn} = (4 \, c \sigma_{\rm T} / 3 \, m_{\rm e} c^2) \, U_{\rm B} \, \gamma^2$, and the magnetic field energy density $U_{\rm B} = B^2 / 8 \pi$, and their escape rate $\dot{\mathcal{N}}_0^{\rm esc}(\epsilon_0) = - \mathcal{N}_0(\epsilon_0) / t_{\rm esc}$.
(iii) The pair injection rate $\dot{\mathcal{N}}_{\rm e}^{\gamma \gamma}(\gamma) = - 4 \, \dot{\mathcal{N}}^{\gamma \gamma}(\epsilon)$ \citep[since $\gamma\approx\epsilon/2$, see][]{cop90}. Their radiative cooling rate, 
by synchrotron emission, $\dot{\mathcal{N}}_{\rm e}^{\rm syn}(\gamma) = - \mathcal{N}_{\rm e}(\gamma) / t_{\rm syn}$ where $t_{\rm syn} = \gamma / |\dot{\gamma}|_{\rm syn}$, and by inverse-Compton scattering $\dot{\mathcal{N}}_{\rm e}^{\rm ic}(\gamma) = - \mathcal{N}_{\rm e}(\gamma) / t_{\rm ic}$ with $t_{\rm ic} = \gamma / |\dot{\gamma}|_{\rm ic}$. both modeled here as catastrophic escape from the analyzed energy range. The inverse-Compton loss term is $|\dot{\gamma}|_{\rm ic} = (4 \, c \sigma_{\rm T} / 3 \, m_{\rm e} c^2) \, U_{\rm 0} \, \gamma^2 \, \eta^{-1}$, where $U_0 = \epsilon_0^2 \, m_{\rm e} c^2 \, \mathcal{N}_0(\epsilon_0)$ is the energy density of the soft photons and the factor $\eta \approx 5^{1.5}$ takes account of Klein-Nishina effects, since  $\gamma \, \epsilon_0 = 1$ and we do not consider the detailed spectral shape \citep[see][]{mod05}.
The influence of inverse-Compton scattering and annihilation on the soft photon field is negligibly small provided $\gamma\gg1$ and $\epsilon\gg1$.

Our system thus has three free parameters, $\epsilon$ (or $B$), $R$ and $L_{\rm inj}$, and three variables, $\mathcal{N}(\epsilon)$, $\mathcal{N}_{0}(\epsilon_0)$ and $\mathcal{N}_{\rm e}(\gamma)$. It is convenient to express the parameters in terms of dimensionless quantities: the high-energy injection and the magnetic compactnesses, 
\eqb
\elli\equiv {\epsilon \, L_{\rm inj} \, \sigma_{\rm T} \over 
4 \pi \, m_{\rm e} c^3 \, R} 
&\textrm{\ and\ }&
\ellb \equiv {\epsilon \, U_{\rm B} \, R \, \sigma_{\rm T} \over 
m_{\rm e} c^2}.
\label{compactness}
\eqe
Note that these quantities contain a factor $\epsilon$ not present in the conventional definitions \citep{gui83}. Instead of the soft photon number density we use the optical depth for photon-photon annihilation
$n_0 \equiv \tau_{\rm \gamma \gamma} = {2 \, \sigma_{\rm T} \, R \mathcal{N}_0(\epsilon_0)/(3\epsilon)}$,
and instead of the hard photon number density, we use the ratio of the hard photon energy density to the energy density of soft photons needed to give $\tau_{\gamma\gamma}=1$:
$n \equiv {U / \left.U_0\right|_{\tau_{\rm \gamma \gamma} = 1}} = {\epsilon^3 \sigma_{\rm T} R\mathcal{N}(\epsilon)/ 6}$. Similarly, for the electron density we use the ratio of the energy density to that of soft photons required for $\tau_{\rm \gamma \gamma} = 1$: 
$n_{\rm e} \equiv {U_{\rm e} /\left.U_0\right|_{\tau_{\rm \gamma \gamma} = 1}} = {\epsilon^3 \, \sigma_{\rm T} \, R \mathcal{N}_{\rm e}(\gamma)/ 24}$,
and measure time in units of the light crossing time $\tau\equiv ct/R$. This leads to the autonomous dynamical system:
\eqb
{\diff \, n \over \diff \tau} &=& - n \, n_0 - n + {2 \over 3} \, \elli 
\nonumber\\
{\diff \, n_0 \over \diff \tau} &=& - n_0 + {1 \over 3} \, \ellb \, n_{\rm e} 
\nonumber\\
{\diff \, n_{\rm e} \over \diff \tau} &=& n \, n_0 - {2 \over 3} \, \ellb \, n_{\rm e} - {4 \over \eta} \, n_{\rm e} \, n_0 \quad .
\label{dynamical}
\eqe

\section{Stationary solutions and their stability}

For all $\elli>0$ and $\ellb>0$ there exists a trivial stationary solution of the system (\ref{dynamical}): $n=2\elli/3$, $n_0=0$, $n_{\rm e}=0$, corresponding to free propagation of the hard photons through a medium in which pairs and soft photons are completely absent. The eigenvalues of the Jacobian matrix evaluated for this solution are all real. For $\elli<3$ they are all negative, so that the solution is stable in this region. This means that the nonlinearity is not sufficient for the system to initiate and sustain a population of pairs if $\elli<3$. Furthermore, in this region, no other physically acceptable ($n_0>0$, $n_{\rm e}>0$) stationary solution exists. For $\elli>3$, at least one of the eigenvalues has a positive real part, so that the solution in which pairs are absent is unstable. However, in this region a second stationary solution emerges, with
$n={2\elli/\left[3\left(1+n_0\right)\right]}$, $n_{\rm e} = {3n_0/\ellb}$ and
$n_0= \left[-6-\eta\ellb +\sqrt{\left(\eta\ellb-6\right)^2+8\eta\ellb\elli}\right]/12$.
Using the Routh-Hurwitz conditions \citep[][17.715]{gradshteynryzhik80}, one
can show that the real parts of the eigenvalues of the Jacobian matrix
corresponding to this physical solution are negative, so that the solution is
stable for all values of $\ellb>0$. All physically acceptable solutions appear
to approach this stable one. We find no chaotic regime. We also have no
physical grounds on which to expect such behavior within the current model.

\begin{figure}
\includegraphics[scale=1.5]{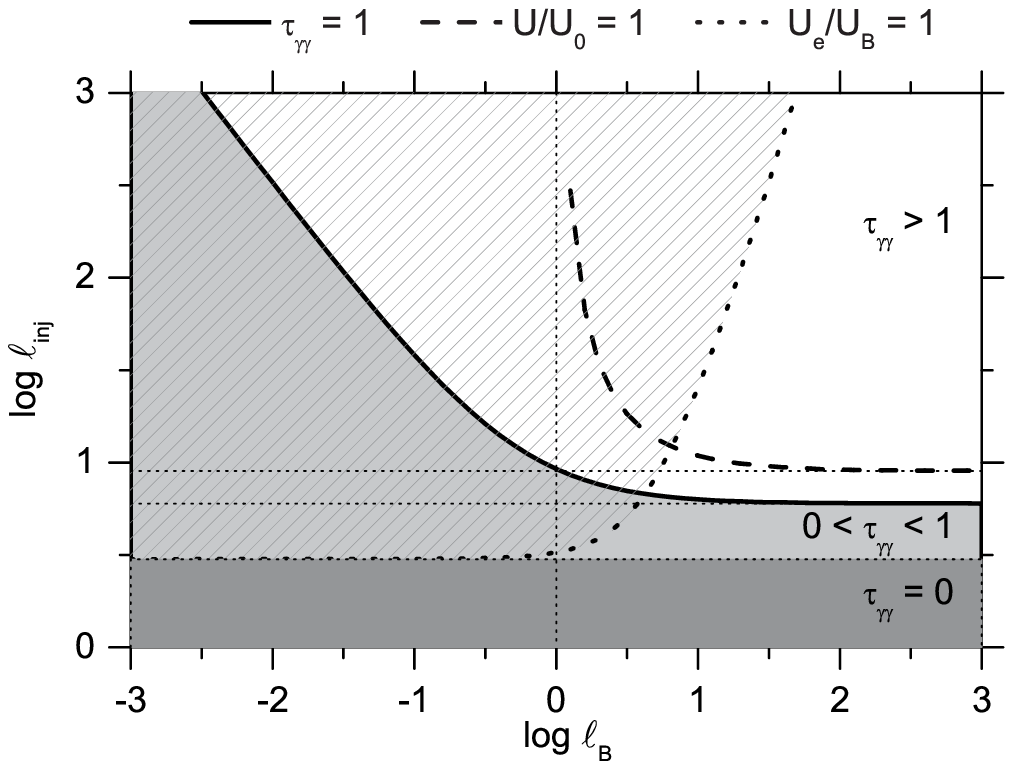}
\caption{%
\label{parameterspace}%
Properties of the stable solutions to the dynamical system of Eq.~(\ref{dynamical}) as functions of the injection and magnetic compactnesses, Eq.~(\ref{compactness}). The dark gray shading denotes solutions with no electrons or soft photons, light gray shading denotes solutions with optical depth to photon-photon absorption less than unity and in the white area strong absorption occurs. The hatched area denotes solutions in which the energy density of the created pairs exceeds that in the magnetic field. The dashed line is the locus of points at which the energy density in soft photons formally reaches equipartition with that in hard photons, the dotted line that of equipartition between pairs and magnetic field.}
\end{figure}

Figure~\ref{parameterspace} illustrates the properties of the stable solutions of this system on the $\ellb$--$\elli$ plane. The horizontal dotted line $\elli=3$ bounds from above the region (shaded dark gray) in which the stable solution contains neither pairs nor soft photons, so that $\tau_{\gamma\gamma} = 0$. The solid line, on which $\elli = 6 + 36/(\eta \, \ellb)$, is the locus of points at which the stable solution has unit optical depth to absorption of high energy photons: $n_0=\tau_{\gamma\gamma}=1$. Strictly speaking, our treatment of the escape probability of hard photons is valid only in the optically thin case $\tau_{\gamma\gamma}<1$, which is shaded light gray. However, for $\elli > 6 + 36/(\eta \, \ellb)$, (white area) we expect on physical grounds that the system tends spontaneously to an optically thick state with almost total absorption of the hard photons. Stable solutions which have equal energy densities of soft and hard photons ($n=n_0$) occur formally when $\elli=3\eta\ellb(3\eta\ellb-12)/ (\eta\ellb-12)^2$, shown as a thick dashed line. This line lies entirely within the optically thick region. Thus, our system is valid only for solutions in which the energy density of soft photons is small compared to that in hard photons. Finally, stable solutions in which the electron and magnetic field energy densities are equal occur when $U_{\rm e}/ U_{\rm B} \equiv 6 \, n_{\rm e}/ \ellb = 1$, shown as a thick dotted line. Solutions which are particle dominated lie above this line in the hatched region of Fig.~\ref{parameterspace}. It is interesting to note that when the optical depth for photon-photon annihilation is less than unity, $\tau_{\gamma \gamma} < 1$, stable solutions can be found that are pair dominated, in the sense that the electron energy density exceeds that in the magnetic field. These solutions occur in the intersection of the hatched and light gray areas of Fig.~\ref{parameterspace}. They imply efficient pair-loading of the system and are particularly relevant for $\ellb \ll 1$, where they occur for all relevant injection compactnesses $\elli>3$.

\section{Applications}

The electrons created in the photon-photon annihilation process cool due to synchrotron and inverse-Compton energy losses. In the analysis presented so far, where a monoenergetic high energy photon field was considered, with fixed photon energy $\epsilon = \epsilon^{\star}$, these losses were modeled as the escape of particles from the relevant energy range. In a more precise treatment, the electrons do not escape catastrophically, but cool (due to synchrotron emission and inverse Compton scattering). Consequently, below the injection energy $\gamma^{\star}=\epsilon^{\star}/2$ the steady state spectrum of the cooled electrons is $N_{\rm e}(\gamma < \gamma^{\star}) \propto \gamma^{-2}$, leading to a synchrotron spectrum at photon energies $\epsilon_0 < \epsilon_0^{\star} = 2 /\epsilon^{\star}$ \citep{kar62}. This corresponds to an emissivity $j_{\epsilon_0 < \epsilon_0^{\star}} \propto {\epsilon_0}^{-1/2}$. In turn these photons absorb high-energy radiation with photon energies $\epsilon > \epsilon^{\star}$. Since the optical depth for the photon-photon annihilation is $\tau_{\rm \gamma \gamma} \propto \epsilon_0 \, \mathcal{N}_0(\epsilon_0) \propto j_{\epsilon_0}$, the condition for $\tau_{\gamma \gamma}>1$ at $\epsilon=\epsilon^{\star}$ ensures automatically $\tau_{\gamma \gamma}>1$ for any $\epsilon > \epsilon^{\star}$. In other words, if the parameters of the considered system are such that it becomes optically thick for the high energy photons with $\epsilon^{\star}$, then all the emission at higher photon energies $\epsilon > \epsilon^{\star}$ will also be absorbed. For an intrinsic luminosity $L_{\rm inj}$ corresponding to the photon energies $\epsilon^{\star}$, the appropriate condition for $\tau_{\gamma \gamma}(\epsilon > \epsilon^{\star}) > 1$ can be therefore expressed as $L_{\rm inj} \geq L_{\rm crit}= \left(10^{-6} {\varepsilon^{\star}_{\rm GeV}}^4 + 4.4 \times 10^5 
R_{\rm pc}/\varepsilon^{\star}_{\rm GeV}\right)\times10^{40}\,\textrm{erg/s}$ .

Supermassive black holes (SMBHs) in active galactic nuclei are an interesting application of the process discussed here. These objects may accelerate protons to high energies \citep{sikoraetal87,bol99}, generating very high energy $\gamma$-ray emission \citep{lev00}. The magnetic field intensity in such systems is high: assuming equipartition between magnetic and accretion-produced radiation gives $B = \left(R / r_{\rm g}\right)^{-1} \, \left(L_{\rm acc} / L_{\rm Edd}\right)^{1/2} \, \left(L_{\rm Edd} / r_{\rm g}^2 c\right)^{1/2}$, where $r_{\rm g} = G \, \mathcal{M} / c^2 \approx 1.5 \times 10^{13} \, \mathcal{M}_8$\,cm is the Schwarzchild radius of the black hole with mass $\mathcal{M} \equiv \mathcal{M}_8 \, 10^8 \, M_{\odot}$, $L_{\rm acc}$ is the accretion luminosity produced within radius $R$, and $L_{\rm Edd} = 4 \pi \, G \, \mathcal{M} \, m_{\rm p} \, c / \sigma_{\rm T} \approx 1.3 \times 10^{46} \, \mathcal{M}_8$\,erg\,s$^{-1}$ is the corresponding Eddington luminosity. SMBHs in AGNs typically have masses $\mathcal{M}_8 \sim (10^{-2}-10^2)$ and luminosities $\lambda \equiv L_{\rm acc} / L_{\rm Edd} \sim (10^{-7} - 1)$. However, lower-mass black holes in the centers of `normal' galaxies, with $\mathcal{M}_8 < 10^{-2}$, also exhibit nuclear activity with very low accretion luminosities $\lambda \sim 10^{-10}$ (e.g., Sgr A$^{\star}$ in the Galactic center). Assuming that the magnetic field scaling holds for all these objects within $R \sim \textrm{few} \times r_{\rm g}$, one finds $B \sim 10^4 \, \left(\lambda / \mathcal{M}_8\right)^{1/2}$\,G, and, hence 
$\left[L_{\rm crit} / 10^{40}\textrm{erg\,s$^{-1}$}\right]$ $\sim 3 \times 10^{-6} \, \left(\lambda / \mathcal{M}_8\right)^{-2/3} + 5 \, \mathcal{M}_8 \, \left(\lambda / \mathcal{M}_8\right)^{1/6}$ at photon energies $\varepsilon^{\star} \sim \left(\lambda / \mathcal{M}_8\right)^{-1/6}$\,GeV.

\begin{figure}
\includegraphics[scale=1.5]{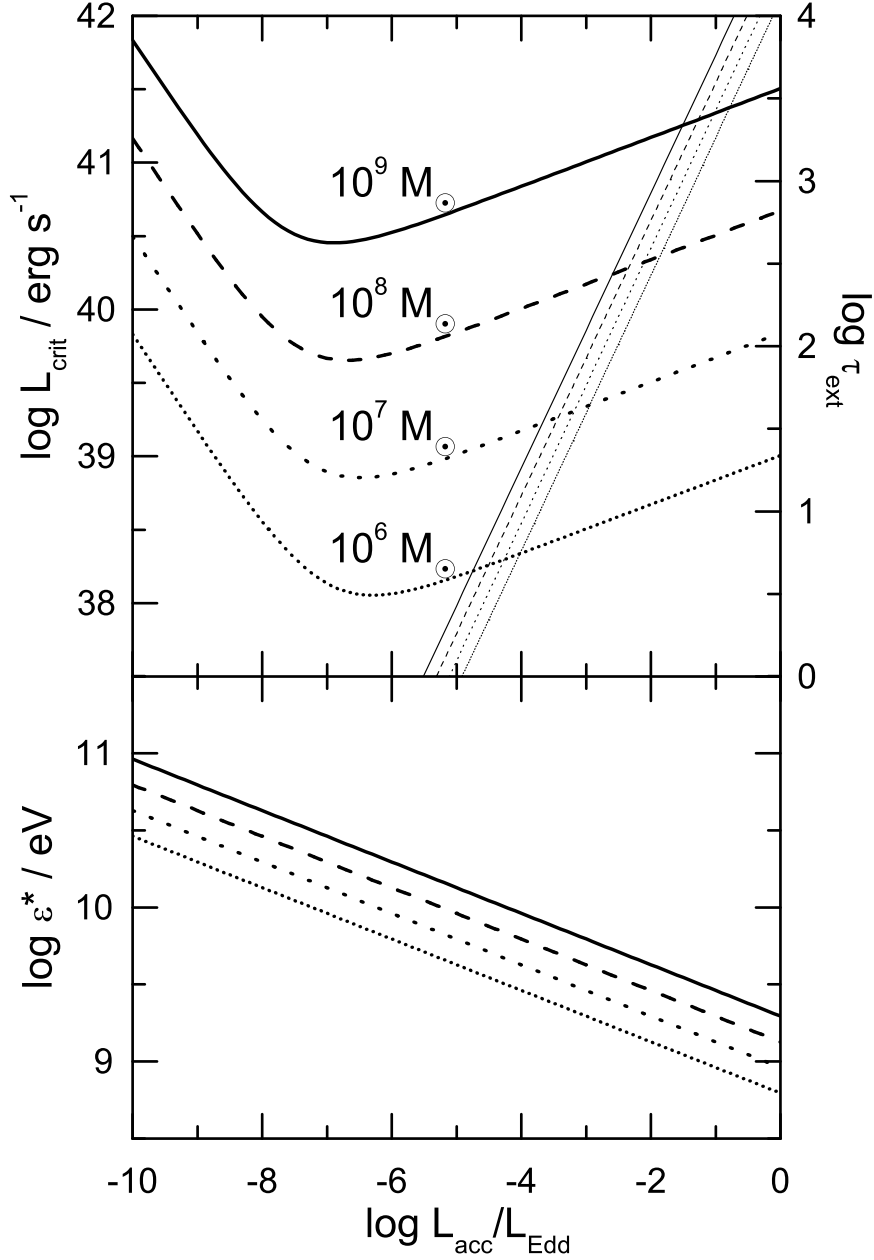}
\caption{%
\label{lbplot}%
Upper panel: the critical injection luminosity $L_{\rm crit}$ that automatically generates a pair density and soft photon field of unit optical depth to photon-photon absorption (thick lines, left-hand ordinate), 
and the opacity of $\varepsilon^{\star}$ photons emitted by accreting matter
(thin lines, right-hand ordinate), both as functions of the accretion luminosity $L_{\rm acc}/L_{\rm Edd}$.
Lower panel: the corresponding maximum gamma-ray energy 
$\varepsilon^{\star}$.
In each case four lines are plotted, corresponding 
to different masses of the SMBH.} 
\end{figure}

The critical luminosity and the corresponding values of $\varepsilon^{\star}$ are plotted in the upper (left-hand ordinate) and lower panels of Figure~\ref{lbplot}, respectively, for different black hole masses. In the case of galactic nuclei hosting the most
massive black holes $\mathcal{M}_8 \geq 1$ no steady (on the light-crossing timescale) high energy $\gamma$-ray emission at photon energies $\varepsilon \geq \varepsilon^{\star} \sim (1-100)$\,GeV can emerge if $\left[\varepsilon^{\star} L_{\varepsilon^{\star}}\right]\geq (10^{41}-10^{42})$\,erg s$^{-1}$. For $\mathcal{M}_8 < 1$, $L_{\rm crit}$ is much lower, $\sim (10^{38}-10^{39})$\,erg s$^{-1}$. This does not mean that lower luminosity sources are able to emit high-energy gamma-rays, since these may also be absorbed on the `external' low-energy radiation of the accreting matter. The relevant optical depth is $\tau_{\rm ext} = (1/3) \, \sigma_{\rm T} \, R \, U_{\rm ext} / \varepsilon_{\rm ext}$, where $U_{\rm ext}$ is the monochromatic energy density of the accretion-related emission at $\varepsilon_{\rm ext} = 2 \, m_{\rm e}^2 c^4 / \varepsilon^{\star}$. A conservative estimate is $U_{\rm ext} = \zeta \, L_{\rm acc} / (4 \pi \, R^2 \, c)$, with $\zeta \approx 0.1$, since, for $\varepsilon^{\star} \sim (1-100)$\,GeV, it is the low-energy photons $\varepsilon_{\rm ext} \sim (0.01-1)$\,keV that are relevant. Using Eq.~(\ref{resonance}) and the assumed scaling of the magnetic field $\tau_{\rm ext} \approx 2.7 \times 10^4 \, \lambda \, \left(\lambda / \mathcal{M}_8\right)^{-1/6}$. The upper panel of Fig.~\ref{lbplot} shows
$\tau_{\rm ext}$ (thin lines, right-hand ordinate), and one sees that automatic quenching of high-energy $\gamma$-ray emission generated in the immediate vicinities of SMBHs competes with the opacity due to emission of the accreting matter only in the case of low accretion rates, $\lambda < 10^{-5}$.

The loading of the system with electron-positron pairs is another interesting aspect of automatic quenching. Defining $\lambda_{\rm inj} \equiv L_{\rm inj}/L_{\rm Edd}$, and again assuming magnetic and radiation pressures are equal, the compactnesses (Eq.~\ref{compactness}) become $\elli \sim 10^6 \, \lambda_{\rm inj} \left(\lambda /\mathcal{M}_8\right)^{-1/6}$ and $\ellb \sim 10^6 \, \lambda \left(\lambda /\mathcal{M}_8\right)^{-1/6}$. Fig.~\ref{parameterspace} shows that the energy density of the pairs substantially exceeds that in the magnetic field if $\elli > 3$, $\ellb<1$, and $\tau_{\rm ext} < 1$.  In the range of $\lambda$ and $\mathcal{M}_8$ of interest for SMBHs one finds $\left(\lambda / \mathcal{M}_8\right)^{-1/6} \sim (1-100)$. Consequently, pair loading is important for those lower luminosity sources, whose gamma-ray output is comparable to, or less than, the accretion luminosity $\lambda \lesssim(10^{-8}-10^{-6}) \lesssim \lambda_{\rm inj}$. 

\section{Conclusions}

The resonance-like behavior of the photon-photon annihilation cross section means that, for magnetic fields between $B = 10^{-9}$\,G and $10^6$\,G, $\gamma$-ray photons between $\varepsilon \approx 0.3$\,GeV and $30$\,TeV (see Eq.~\ref{resonance}) can be absorbed on the synchrotron radiation of the created pairs. This energy range is now, or will shortly, be accessible to the {\it AGILE} and {\it GLAST} satellites, together with the Cherenkov Telescopes {\it HESS}, {\it MAGIC}, {\it CANGAROO}, and {\it VERITAS}. The nonlinear dynamics of these $\gamma$-ray sources render them unstable to runaway pair production when the injection compactness $\ell_{\rm inj} > 3$ (see Eq.~\ref{compactness}). The spontaneously produced optical depth to absorption saturates at value greater than unity for photons with energies above $\varepsilon$, if $\ell_{\rm inj} \geq 6 + 36/(\eta \, \ellb)$, where $\eta \approx 5^{1.5}$ and $\ellb$ is the magnetic compactness
(see Eq.~\ref{compactness}). In this case the source automatically quenches its own $\gamma$-ray emission. As a corollary, we find that systems with small $\ellb < 1$ but large $\ell_{\rm inj} > 3$ suffer significant pair loading. This implies that $\gamma$-ray sources in the vicinity of low accretion-rate supermassive black holes can be automatically quenched and their jets loaded with electron-positron pairs. 

\acknowledgments

\L .S. was supported by MEiN through the project 1-P03D-003-29 in years 2005-2008.


\clearpage

\clearpage

\end{document}